\documentclass[pra,aps, a4,twocolumn,amsmath,amsfonts,amssymb,superscriptaddress,epsfig]{revtex4}
\usepackage{psfrag}
\usepackage{epsfig}
\usepackage{color} 
\usepackage{graphicx}
\def\bra#1{\langle{#1}|}
\def\ket#1{|{#1}\rangle}
\def\pro#1{|{#1}\rangle\langle{#1}|}
\def\id{{\openone}}
\newcommand{\PP}{\ensuremath{\mathcal{P}}}

\usepackage{multirow}
\usepackage{calc}
\usepackage{boxedminipage}
\usepackage{ifthen}
\usepackage{dsfont}
\usepackage{pifont}
\usepackage{bbm}
\usepackage{graphicx}
\usepackage{makeidx}
\usepackage{amsfonts}
\newcommand{\field}[1]{\mathbb{#1}}

\newcommand{\N}{\field{N}}
\newcommand{\eins}{\openone}

\renewcommand{\descriptionlabel}[1]%
        {\hspace{\labelsep}\textsf{#1}}

\newcommand{\Mentrylabel}[1]%
{\raisebox{Opt}[1ex][Opt]{\makebox[labelwidth][1]%
 {\parbox[t]{\labelwidth}{\hspace{Opt}\textsf{#1:}}}}}
{\begin{entry}}%
{\end{entry}}
\def\tr{\mbox{tr}}

\def\refe#1{(\ref{#1})}

\def\N2{\mbox{N}_2}

\newsavebox{\tempbox}

\begin{document}
\title {Parameter estimation for mixed states from a single copy
}
\author{}
\date{\today
}
\begin{abstract}
Given a single copy of a mixed state of the form
$\rho=\lambda\rho_1+(1-\lambda)\rho_2,$ what is the 
optimal measurement to estimate the parameter 
$\lambda$, if $\rho_1$ and $\rho_2$ are known?
We present a general strategy to obtain the optimal 
measurements employing a Bayesian estimator. 
The measurements are chosen to minimize the deviation 
between the estimated- and the true value of $\lambda$. 
We explicitly determine the optimal measurements for 
a general two-dimensional system and for important higher
dimensional cases.
\end{abstract}

\author{Thomas Konrad}
\affiliation{School of Physics, University of 
KwaZulu-Natal, 
Private Bag 54001, Durban 4000, South Africa
}

\author{Otfried G\"uhne}
\affiliation{Institut f\"ur Quantenoptik und Quanteninformation,
\"Osterreichische Akademie der Wissenschaften,
A-6020 Innsbruck, Austria}

\author{J\"urgen Audretsch}
\affiliation{Fachbereich Physik, Universit\"at Konstanz, Fach M 674, 
D-78457 Konstanz, Germany}

\author{Hans J. Briegel}
\affiliation{Institut f\"ur Quantenoptik und Quanteninformation,
\"Osterreichische Akademie der Wissenschaften,
A-6020 Innsbruck, Austria}
\affiliation{Institut f\"ur Theoretische Physik, Universit\"at Innsbruck, 
Technikerstra{\ss}e 25, A-6020 Innsbruck, Austria}

\maketitle

\section{Introduction}
\label{sec:intro}

The estimation of quantum states is one of the 
basic primitives in quantum theory. This general 
task appears in several modifications and situations. 
When many copies of the state are available, one may 
try to obtain maximal information on the quantum state 
by {employing} state tomography \cite{estimation}, although 
the effort may become overwhelming, if the dimension 
of the system  increases. Furthermore, tomography is clearly 
not a viable way, if only few, or even only a single copy 
of the state are available. 

Nevertheless, if some {\it a priori} information about the 
state is given,  there are several estimation problems which 
can be  meaningfully posed and solved even on the single copy
level. The most prominent is state discrimination, where one 
knows that the {system  is in one of several given states}, and 
one  needs to decide in which one. This problem has been discussed
in several variations, either one aims {for} an unambiguous
discrimination, or for a discrimination with a minimal error probability 
\cite{estimation, estimation2}. Moreover, programmable 
state discriminators have been proposed \cite{bergou}. 
Other problems considered are state estimation from several 
copies with collective or separable measurements \cite{munos}, 
the estimation of certain state parameters (like the time in an unitary 
evolution) \cite{braunstein}, as well as estimation of the state 
after a generalized measurement when the premeasurement state is 
unknown \cite{audretschdiosikonrad03}. 

In this paper, we consider the following problem: Let us assume
that an apparatus is given, which produces for a given input 
$\lambda$ the state
\begin{equation}
\label{eins}
\rho(\lambda) =\lambda\rho_1+(1-\lambda)\rho_2\,,
\end{equation}
where $\rho_1$ and $\rho_2$ are known. Let us further assume 
that $\lambda$ is determined by some well characterized random 
number generator, but we don't know its actual value. The task 
is now to estimate $\lambda$ from a single copy of $\rho(\lambda).$ 

Such estimation problems can occur in several realistic 
situations. For instance, one may consider some process of 
decoherence, where the single copy of $\rho(\lambda)$
{describes the state of a single atom coupled to its environment, 
and the task is to estimate the rate of 
decoherence. In this case $\rho_1$ represents an initial 
state, while $\rho_2$ is the state of thermal equilibrium, which 
the system assumes eventually.} Another 
example may be that $\rho(\lambda)$ is the reduced state of 
some multipartite pure state $\ket{\psi(\lambda)}$, where one 
tries to estimate $\lambda.$ We will discuss the {decoherence} example 
later in more detail.

For the estimation one has to perform some measurement.
The must general measurement is described by a positive 
operator valued measure (POVM) and we ask for the POVM 
that minimizes the {expected} deviation between the 
true value of $\lambda$ and the estimated one. 
{Here we take the viewpoint that the value of
$\lambda$ is not exactly known  beforehand but we may have
information about it in form of a prior probability
distribution. After a measurement this prior probability
distribution can be updated according to Bayes
theorem. Minimizing the deviation of the estimate from  
$\lambda$ expected from the posterior
probability distribution leads to the so-called Bayesian
estimator {(cp.\ for example \cite{LeonardHsu01})}. 

Bayesian estimators have been successfully applied in state
discrimination problems. They are also appropriate in situations 
where a
sequence of consecutive measurements is carried out and the
estimate has to be updated after each measurement. This is for
example the case in real-time monitoring of the dynamics of quantum
systems. Using a Bayesian estimator it is possible to monitor
oscillations of single qubits in real-time with high accuracy  
by means
of a sequence of weak measurements \cite{audretsch, audretsch2}.
{The Bayesian estimator is also intimately related to the 
continuous estimation of the wave function of a system with 
arbitrary-dimensional
Hilbert space by means of continuous measurement \cite{audretsch3}.} 
Although the search for the best measurement in connection with an
Bayesian estimator is performed in this article for the particular
task to estimate the parameter $\lambda$ in Eqn.\ (\ref{eins}), it may
be carried out in an analogous way in different set-ups.}

For the case when $\rho$ is a two-level system, i.e., a qubit, 
we determine the optimal measurement strategy for arbitrary 
$\rho_1$ and $\rho_2.$ Interestingly, 
{for the case of completely unknown $\lambda$}, it turns out that if 
$\rho_1$ and $\rho_2$ do not have the same purity, the optimal 
measurement does not commute with $\rho_1 - \rho_2,$  and 
hence does not coincide with the intuitive guess which one would 
make from the Bloch sphere picture.
Finally, we discuss how the results from the qubit case can 
readily be used to  solve this problem  for important higher 
dimensional cases. 

As already mentioned, similar parameter estimation problems have 
in generality been studied in Ref.~\cite{braunstein} and general 
conditions on the information which can be obtained about the 
parameter have been formulated. Our aim, however, is to explicitly 
construct the optimal measurements. The knowledge of these explicit 
measurements may further be used to improve schemes for the 
observation of oscillations with small disturbance \cite{audretsch,audretsch2}.

Our paper is organized as follows: In Section II we pose the problem in
mathematical terms and derive a general condition on the optimal POVM.
In Section III we show how this general condition can already reduce the 
set of POVMs which have to be considered. In Section IV we solve the problem
for qubits. We first derive the optimal measurement, if $\rho_1$ and $\rho_2$
are pure, then we consider the general case. We also discuss how one can apply 
the results for estimating decoherence rates. Finally, in Section 
V we derive some results for higher dimensional systems.

\section{Bayesian estimator and condition for an optimal measurement}

In this paper, we consider the following task. Let us assume 
that we have a single quantum system with $d$-dimensional 
Hilbert space $\mathcal H$. {Its} state is known to be a 
mixture of two states $\rho_1$ and $\rho_2$, i.e.,
\begin{equation}\label{rholambda}
\rho_\lambda=\lambda\rho_1+(1-\lambda)\rho_2\,, 
\end{equation}
where for the parameter $\lambda$ only some probability 
distribution is known. What is the measurement which leads 
to an optimal estimation of the unknown parameter $\lambda$? 

The estimation of $\lambda$ has to proceed in two steps: 
First, a measurement is carried out on the system. The 
statistics of a general measurement can be characterized 
by a POVM and throughout this paper we denote by $\PP=\{E_m\}$ 
a POVM with the effects $E_m.$ In a second step the parameter 
$\lambda$ is estimated by a number $g_m$ (the estimate) which 
depends on the result of the measurement $m$.

Let us first discuss the construction of the optimal 
estimate $g_m$ for a given outcome $m$. A criterion for 
the optimality of the estimate $g_m$ can be formulated as 
follows: $g_m$ is optimal iff the cost function
\begin{equation}
c(\lambda,g_m)=(\lambda-g_m)^2
\end{equation}
is expected to assume its minimum:
\begin{equation}
{\mathcal E}_m\left((\lambda-g_m)^2\right) 
:= 
\int (\lambda-g_m)^2 p(\lambda| m)d\lambda 
\stackrel{!}{=} 
\min.
\end{equation}
Here  the expectation value $\mathcal E$ is taken with 
respect to the posterior probability distribution of 
$\lambda$, which includes the information contained 
in the occurrence of the measurement result $m$:
\begin{equation}
\label{posterior}
p(\lambda|m)=\frac{p(m|\lambda)p(\lambda)}{p(m)}\,,
\end{equation}
where $p(m)=\int p(m|\lambda)p(\lambda)d\lambda$ is 
the probability to obtain the measurement result $m$ 
averaged over all possible occurring states. 

Taking into account the linearity of the expectation value, 
one can directly verify that the optimal estimator $g_m$ 
is equal to the expected value of $\lambda$:
\begin{eqnarray}
{\mathcal E}_m\left((\lambda-g_m)^2\right)
&=& {\mathcal E}_m\left((\lambda)^2\right)-
  2g_m{\mathcal E}_m(\lambda)+g_m^2\,,
\nonumber\\ 
&=& \left({\mathcal E}_m(\lambda)-g_m\right)^2 +
\mbox{Var}_m(\lambda)\,,\label{bla}
\end{eqnarray}
where 
\begin{equation}\label{Var}
\mbox{Var}_m(\lambda)={\mathcal E}_m(\lambda^2)-{\mathcal E}_m(\lambda)^2
\end{equation}
represents the variance of $\lambda$. The right-hand side of
(\ref{bla}) assumes a minimum for $g_m={\mathcal E}_m(\lambda)$. 
Such a value $g_m$ is also called a Bayesian estimate. 

Having derived the optimal estimate $g_m$ for a certain 
outcome $m$ we can consider the optimal choice of the POVM. 
Now, the optimal measurement is the one which leads to the 
smallest expected costs averaged over all outcomes $m$:
\begin{equation}
\sum_m p(m) {\mathcal E}_m\left((\lambda-g_m)^2\right)
\stackrel{!}{=} 
\min.
\end{equation}
Hence the effects $\{E_m\}$ have to minimize the mean variance 
$\sum_m p(m) \mbox{Var}_m(\lambda)$. Note that this variance, 
just as the expectation of the costs, is defined with respect 
to the posterior probability density  $p(\lambda|m)$. 

For the sake of simplicity we choose in the following an 
equally distributed prior density $p(\lambda)=1$. We will 
see later in an example, that this {imposes} not a big restriction. 
Using 
$p(m|\lambda)= \tr[E_m \rho_\lambda]$ and the probability 
to obtain measurement result $m$ 
\begin{align}
p(m) 
&=  \int_0^1\, p(m|\lambda)p(\lambda) d\lambda 
\nonumber\\
&= \int_0^1\,\tr[E_m\left(\lambda\rho_1 + (1-\lambda)\rho_2\right)] d\lambda
\nonumber\\
&= \tr\left[E_m\frac{1}{2}(\rho_1+\rho_2)\right]
\label{pm}
\end{align}
we obtain for the posterior probability density  in 
Eq.~(\ref{posterior}) 
\begin{equation}
p(\lambda|m) 
= 
\frac{\tr[E_m\rho_\lambda]}{\tr\left[E_m\frac{1}{2}(\rho_1+\rho_2)\right]}\,.
\label{pm2}
\end{equation}
In order to minimize  the mean variance of $\lambda$ 
we have to first compute the expectation value of $\lambda$ 
and $\lambda^2$ with respect to the posterior probability 
density:
\begin{align}
{\mathcal E}_m(\lambda)
&= \int_0^1\,\lambda \,p(\lambda|m) d\lambda 
= \frac{\int_0^1\, \lambda\,\tr[E_m\rho_\lambda]d\lambda}
{\tr\left[E_m\frac{1}{2}(\rho_1+\rho_2)\right]}
\nonumber\\  
&= \frac{\tr\left[E_m\left(\frac{2}{3}\rho_1 + 
\frac{1}{3}\rho_2\right)\right]}
{2\,\tr\left[E_m\frac{1}{2}(\rho_1+\rho_2)\right]} 
\label{expect}
\end{align}
and in a similar calculation
\begin{equation}
{\mathcal E}_m(\lambda^2)
= \int_0^1\,\lambda^2 \,p(\lambda|m)
d\lambda\nonumber
= \frac{\tr\left[E_m\left(\frac{3}{4}\rho_1 + 
\frac{1}{4}\rho_2\right)\right]}{3\,\tr\left[E_m\frac{1}{2}(\rho_1+\rho_2)\right]}.
\label{pm3}
\end{equation}
The mean variance thus amounts to (see Eqns.~(\ref{Var}, \ref{pm})):
\begin{equation}
\label{vari}
\sum_m p(m) \mbox{Var}_m(\lambda) 
=  \frac{1}{3}- \frac{1}{4}
\sum_m \frac{\left(\tr\left[E_m\left(\frac{2}{3}\rho_1 + 
\frac{1}{3}\rho_2\right)\right]\right)^2}
{\tr\left[E_m\frac{1}{2}\left(\rho_1+ \rho_2\right)\right]}\,.
\end{equation}
We can thus summarize the main result of this section: 
\\
{\bf Proposition 1.} The optimal POVM $\PP = \{E_m\}$ is the 
one which maximizes
\begin{equation}
\label{QvonP}
Q(\PP):=
\frac{1}{4}
\sum_{m} \frac{\left(\tr\left[E_m\left(\frac{2}{3}\rho_1+ \frac{1}{3}\rho_2\right)\right]\right)^2}
{\tr\left[E_m\frac{1}{2}\left(\rho_1+ \rho_2\right)\right]}. 
\end{equation}
 
Note that using the fact that $\frac{2}{3}\rho_1+ \frac{1}{3}\rho_2 
=\frac{1}{2}(\rho_1+ \rho_2) + \frac{1}{6}(\rho_1- \rho_2)$ the 
quantity $Q(\PP)$ may be rewritten as 
\begin{equation}
\label{QvonPneu}
Q(\PP):=
\frac{1}{4}
\Big( 1+ 
\sum_{m} \frac{\left(\tr\left[E_m\left(\rho_1-\rho_2\right)\right]\right)^2}
{3 \tr\left[E_m\left(\rho_1+ \rho_2\right)\right]}
\Big).
\end{equation}
which is manifestly invariant under the permutation of $\rho_1$ and $\rho_2.$
Further note that Eq.~\refe{expect} allows us to calculate the 
Bayesian estimate $g_m$ of $\lambda$, given a measurement outcome 
associated with the effect operator $E_m$, since 
$g_m={\mathcal E}_m(\lambda)$.  

{The condition for an optimal measurement in case of 
a general prior probability distribution $p(\lambda)$ 
can be determined from the corresponding mean variance, 
which is of similar form as (\ref{vari}). It can be derived 
as in Eqs.~(\ref{pm}, \ref{pm2}, \ref{expect}, \ref{pm3}) 
with a general $p(\lambda)$ and reads:
\begin{align}
\label{varigen}
& \sum_m p(m) \mbox{Var}_m(\lambda) \nonumber\\
& =  \overline{\lambda^2}- (\bar{\lambda})^2
\sum_m \frac{\left(\tr\left[E_m\left(\frac{ \overline{\lambda^2}}{\bar{\lambda}}\rho_1 + 
\left(1-\frac{ \overline{\lambda^2}}{\bar{\lambda}}\right)\rho_2\right)\right]\right)^2}
{\tr\left[E_m\left(\bar{\lambda}\rho_1+ \left(1-\bar{\lambda}\right)\rho_2\right)\right]}\,,
\end{align}
where $\overline{\lambda^n}$ is the $n$th moment of $p(\lambda)$, i.e.,
$\overline{\lambda^n}= \int_0^1 \lambda^n p(\lambda)d\lambda$. 
Again, this may be rewritten in a way similar to Eq.~(\ref{QvonPneu}) which is 
invariant under permutation of $\rho_1$ and $\rho_2.$ Although we
consider in the following an equidistributed $\lambda$,
thus $p(\lambda)=1$, we illustrate by an example in section \ref{four}
that the results we obtain can readily be transcribed for the general case.}

\section{Properties of optimal measurements}
\label{condopt}

Now we show two facts about the POVM which minimizes the 
expression of Eq.~(\ref{vari}): We first show that this POVM
can be chosen to have pure effects, {i.e.\ effects which are of
rank one}. Then, we show that this  POVM must be a so-called 
extremal POVM.

To show that the POVM which minimizes the expression of 
Eq.~(\ref{vari}) has pure effects $E_i,$ assume that we have 
a POVM where one effect, say $E_1,$ is not pure. Then we can 
decompose $E_1$ as $E_1= E_1^A + E_1^B$ where $E_1^{A}$ and 
$E_1^{B}$  are linearly independent positive operators, which
can serve as new effects. The new POVM with the effects 
$E_1^A, E_1^B, E_2, E_3,...$ gives the same or a smaller  
value for the cost function. Indeed, from Eq.~(\ref{vari}) 
it follows that it suffices to show
\begin{equation}
\frac{(Z_1)^2}{N_1}
\leq
\frac{(Z_1^A)^2}{N_1^A}
+
\frac{(Z_1^B)^2}{N_1^B}
\label{bedingung1}
\end{equation}
where 
$Z_\beta^\alpha = 
\tr[E_\beta^\alpha(\frac{2}{3}\rho_1+\frac{1}{3}\rho_2)]$
and 
$N_\beta^\alpha=\tr[E_\beta^\alpha
\frac{1}{2}(\rho_1+ \rho_2)].$
Eq.~(\ref{bedingung1}) can be straightforwardly verified, 
using the facts that $Z_1=Z_1^A+Z_1^B$ and $N_1=N_1^A+N_1^B.$
Thus, for our search for optimal measurement strategies, it 
suffices to consider POVMs with pure effects.

There are further constraints on the optimal POVM $\PP.$ 
These follow from the fact that the set of all POVMs is 
a convex set. Given two POVMs $\PP^{(1)}=\{E_m^{(1)}\}$ 
and $\PP^{(2)}=\{E_m^{(2)}\}$ with $K$ outcomes, the convex 
combination $\PP = p\PP^{(1)}+(1-p)\PP^{(2)}=\{pE_m^{(1)}+(1-p)E_m^{(2)}\}$
is again a POVM with $K$ outcomes. {On the other hand} 
there are POVMs 
which can not be expressed as a convex combination of two different 
POVMs, these are called {\it extremal} \cite{dariano}. 
For our purpose, it is important to note that for convex 
combinations 
\begin{equation}
Q(\PP) \leq p Q(\PP^{(1)}) + (1-p) Q(\PP^{(2)})\label{beding2}
\end{equation}
holds. This follows because it is true for each of the 
summands of $Q$. This can be verified e.g. for the summand 
with the  effect $E_1 = p E^{(1)}_1+(1-p)E^{(2)}_1$ by 
replacing the effects $E_1^A$ and $E_1^B$ in inequality 
(\ref{bedingung1})  by $p  E^{(1)}_1$ and
$(1-p)E^{(2)}_1$, respectively. Since a generic convex 
combination on the right hand side of (\ref{beding2}) 
assumes only values less than
 $\mbox{Max}\{Q(\PP^{(1)}), Q(\PP^{(2)})\}$, only extremal 
POVMs can maximize $Q$
and thus minimize the expected costs. We can summarize: 
\\
{\bf Proposition 2.} 
When maximizing  $Q$ in Eq.~(\ref{QvonP}) it suffices to 
consider extremal POVMs with pure effects.

Indeed, we can require both conditions at the same time, 
since POVMs with pure effects, which are not extremal, 
can be written as a convex combination of extremal POVMs 
with pure effects. The two conditions on the POVMs are 
a priori independent. However, for special cases, {certain
relations} are known. For instance, it has been shown that 
all extremal POVMs for qubits have pure effects and maximally 
four outcomes \cite{dariano}. This fact we will exploit in 
the following. 

\section{Optimal POVM for qubits}
\label{four}
In this section we consider the problem for qubits, that 
is,  $\rho_1$  and $\rho_2$ are qubit states. The 
representation of these states 
on the Bloch sphere will enable us to determine the optimal POVM. 
We will first derive some general formulation of the optimization
problem in terms of Bloch vectors. Then we will solve the problem 
for the case where $\rho_1$ and $\rho_2$ are pure and finally 
for general  $\rho_1$ and $\rho_2.$

We know already that the effects $E_m$ of an optimal measurement 
are pure, i.e.,
\begin{equation}
E_m= \alpha_m\pro{0_m}.
\end{equation}
Here, the states $\ket{0_m}$ span the whole Hilbert space 
but they are in general not {mutually} orthogonal. Since we are dealing with 
qubits the projectors $\ket{0_m}\bra{0_m}$ can be expressed by means 
of Bloch vectors $\vec{r}_m=\bra{0_m}\vec{\sigma}\ket{0_m}$:
\begin{equation}
\pro{0_m} = \frac{1}{2}(\id + \vec{r}_m \cdot \vec{\sigma})\,,
\end{equation}
where the components of the vector $\vec{\sigma}$ are the 
Pauli operators $\sigma_x$,  $\sigma_y$ and $\sigma_z$. Thus 
the effects read
\begin{equation}\label{para}
E_m = p_m(\id + \vec{r}_m \cdot \vec{\sigma})\,
\end{equation}
with $p_m:=\alpha_m/2$. From the completeness of the effects 
$\sum_m E_m=\id$ we obtain the following constraints for the 
weights $p_m$ and the vectors $\vec{r}_m$:
\begin{equation}\label{constraints}
\sum_m p_m =1 \quad \mbox{and}\quad \sum_m p_m\vec{r}_m =\vec{0}
\end{equation}

The optimal POVM maximizes $Q$ in Eq.~(\ref{QvonP}). Expressing the 
states $\rho_a := (2\rho_1+ \rho_2)/3$ and $\rho_b := (\rho_1+ \rho_2)/2$ 
in terms of Bloch vectors $\vec{r}_i=\tr[\vec{\sigma}\rho_i]$ 
\begin{align}
\rho_a =\frac{1}{2}(\id +\vec{r}_a \cdot \vec{\sigma})\,,
& \;\;\;\;
\rho_b =\frac{1}{2}(\id + \vec{r}_b \cdot \vec{\sigma})\,,
\end{align}
$Q$ reads 
\begin{equation}
Q=\frac{1}{4}\sum_{m}p_m \frac{(1+\vec{r}_m \cdot \vec{r}_a)^2}
{1+\vec{r}_m\cdot \vec{r}_b}\,.
\label{Qqubita}
\end{equation}
The quantity $Q$ can be further simplified by expanding it in powers 
of  the difference vector $\Delta \vec{r}= \vec{r}_a-\vec{r}_b$,
\begin{align}
\label{Qqubit}
Q 
& = 
\frac{1}{4}\sum_{m}p_m 
\frac{\left( 1+\vec{r}_m \cdot \left(\vec{r}_b+\Delta \vec{r}\right)\right)^2}
{1+\vec{r}_m\cdot\vec{r}_b}
\nonumber\\
& = \frac{1}{4}\sum_{m}p_m\left(1 +\vec{r}_m \cdot\left(\vec{r}_b+2\Delta \vec{r}\right) +
\frac{\left(\vec{r}_m\cdot\Delta \vec{r}\right)^2}
{1+\vec{r}_m\cdot\vec{r}_b}\right)
\nonumber\\
& = \frac{1}{4} \left( 1+\sum_{m}p_m\frac{\left(\vec{r}_m\cdot\Delta \vec{r}\right)^2}
{1+\vec{r}_m\cdot\vec{r}_b}\right)\,,
\end{align}
where we have employed conditions \refe{constraints} to obtain 
the last line from the second. Starting with this representation 
of $Q$, we can further reduce the set of interesting POVMs for 
our task. 

{\bf Proposition 3.} When determining the optimal POVM for qubits 
it suffices to consider POVMs with pure effects and maximally 
three outcomes where the vectors $\vec{r}_m$ lie in the plane 
spanned by $\vec{r}_a, \vec{r}_b$ and $\vec{0}$ in the Bloch 
sphere. 

{\it Proof.} 
In fact the effects $E_m$ of any qubit-POVM $\PP$ can be
represented by means of Bloch vectors $ \vec{r}_m$:
\begin{equation}
E_m=p_m(\id+\vec{r}_m \cdot \vec{\sigma})
\end{equation} 
where the $p_m\,,\,\vec{r}_m$ satisfy condition
(\ref{constraints}). Positivity of $E_m$ implies $0\le \|
\vec{r}_m\|\le 1$ with $\|\vec{r}_m\|=1$ iff $E_m$ is pure.
  
Now, let us assume without {restriction of} generality 
that the plane spanned by $\vec{r}_a, \vec{r}_b$ and 
$\vec{0}$ is the {$x$-$z$ plane}. Starting from the vectors 
$\vec{r}_m$ which describe the effects $E_m$ of the POVM 
$\PP$ we consider their projection 
$\vec{q}_m = \vec{r}_m\vert_{xz}$ on the  $x$-$z$ plane. 
This results in a new POVM $\tilde{\PP}$ with the effects
\begin{equation}
\tilde{E}_m=p_m(\id+\vec{q}_m \cdot \vec{\sigma})
\end{equation}
This POVM {possesses in general not only} pure effects, since the
$\vec{q}_m$ are {in the generic case} not normalized. However, 
Eqs.~(\ref{Qqubita}, \ref{Qqubit}) are still valid, which 
implies that $Q(\PP)=Q(\tilde{\PP}).$ To proceed, we can now 
decompose the non-pure effects $\tilde{E}_m$ into pure ones. 
This can for example be done by means of spectral decomposition:
\begin{align}
\tilde{E}_m
 &= p^A_m \frac{\id + \hat{\vec{q}}_m \cdot \vec{\sigma}}{2}+
 p^B_m\frac{\id - \hat{\vec{q}}_m \cdot \vec{\sigma}}{2}\\
 &=: \tilde{E}^A_m + \tilde{E}^B_m\,,\nonumber
\end{align}
where the eigenvalues are given by 
$p^{\mbox{\tiny $A/B$}}_m=p_m(1\pm \|\vec{q}_m\|)$.
Please observe, that the Bloch vectors $\pm\hat{\vec{q}}_m:=\pm
\vec{q}_m/\|\vec{q}_m\|$ of the pure effects 
$\tilde{E}^{\mbox{\tiny $A/B$}}_m$ point into the $x$-$z$ plane. 
Thus it suffices to consider POVMs with pure
effects (but maybe many outcomes) in the $x$-$z$ plane. Finally, 
we can choose extremal POVMs. If the POVM with many pure effects
in the  $x$-$z$ plane is not extremal, we can write it as a convex
combination of extremal ones, which obviously have to have pure 
effects 
in the $x$-$z$ plane again. The point is, that it is known that
for extremal POVMs with four outcomes the projectors cannot lie 
in one plane \cite{dariano}. This proves the claim.
$\hfill \Box$

In the following we study the problem of optimal measurements
beginning with the simplest case, the optimal von Neumann
measurement for pure states $\rho_1$ and  $\rho_2$. Then we attempt to
show that this von Neumann measurement leads to the least mean
variance compared {to} all generalized qubit measurements. 
We increase the level of generality by studying the case of 
mixed states $\rho_1$ and
$\rho_2$.  

\subsection{Optimal PVM for mixtures of pure qubit states}

Let us { first} consider pure states $\rho_1$ and  $\rho_2$ of a 
qubit, and find the optimal  von Neumann  measurement, i.e., 
the PVM $\{P_+=\pro{\psi},P_-=\id-\pro{\psi}\}$ for which 
the expected variance of $\lambda$ is least. The projectors $\{P_+,P_-\}$  
can be expressed by means of normalized Bloch vectors
$\vec{r}_+=\bra{\psi}\vec{\sigma}\ket{\psi}\equiv\vec{r}$ 
and $\vec{r}_-=-\vec{r}$, 
\begin{equation}
P_\pm= \frac{1}{2}(\id \pm \vec{r}\vec{\sigma})\,.
\end{equation}
This implies that the weights which appear in the parameterization 
of the pure effects (\ref{para}) are given by $p_\pm=1/2$.

\begin{figure}
\psfrag{r1}{$\vec{r}_1$}
\psfrag{r2}{$\vec{r}_2$}
\psfrag{r12}{$\vec{r}_1-\vec{r}_2$}
\psfrag{ra}{$\vec{r}_a$}
\psfrag{rb}{$\vec{r}_b$}
\psfrag{dr}{$\Delta\vec{r}$}
\psfrag{rr}{$\vec{r}$}
  \begin{center}
   \epsfig{file=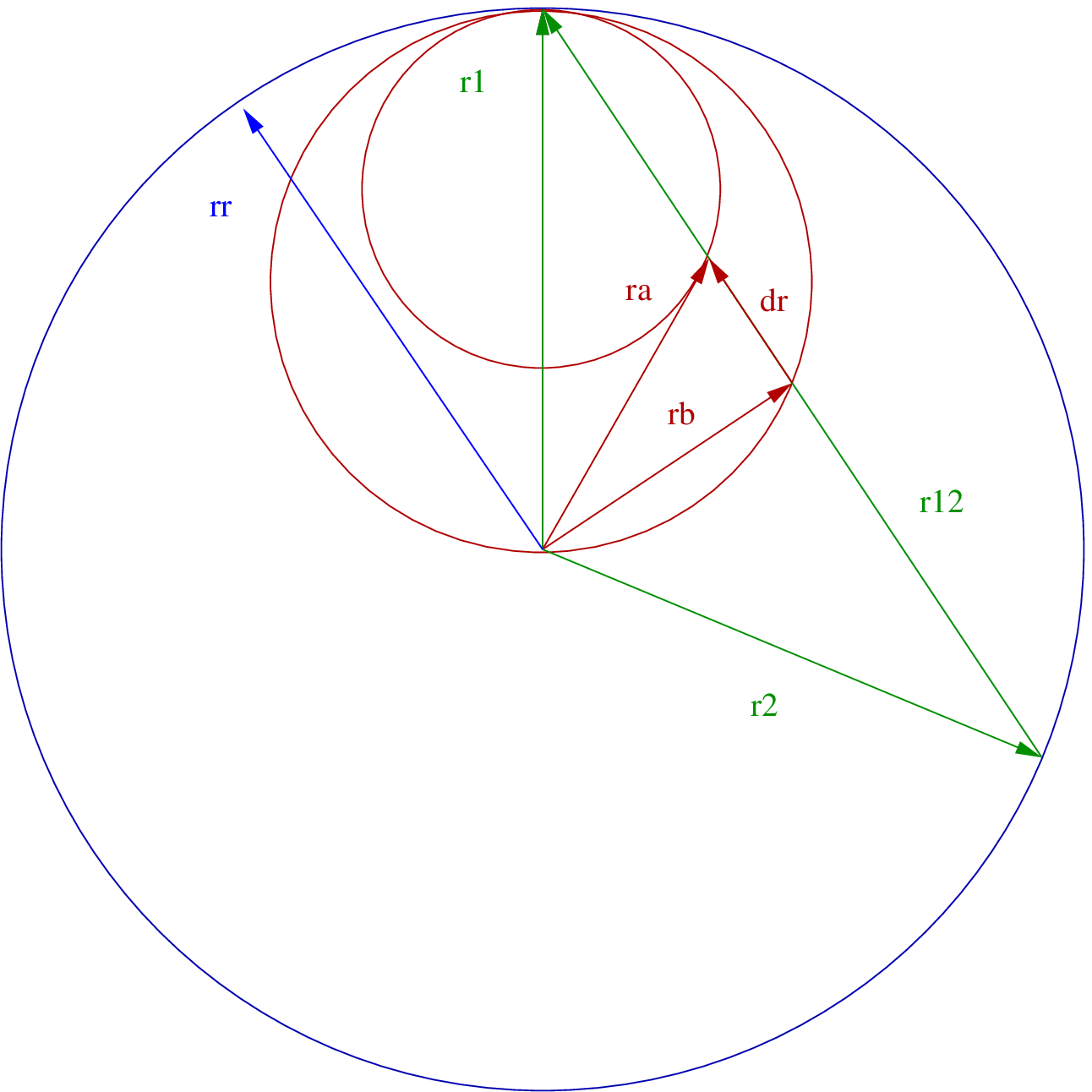,width=0.8\linewidth}
  \end{center}
\caption{Schematic figure of the Bloch representation for the case  
where $\rho_1$ and $\rho_2$ are pure. The Bloch vectors $\vec{r}_b$ 
and $\Delta \vec{r}\propto\vec{r}_1-\vec{r}_2$ are orthogonal, because of 
the circle of Thales. $\vec r$ denotes the direction of the considered 
measurement. In the calculation, it turns our that for the optimal measurement
 $\vec r$ is parallel to $\vec{r}_1-\vec{r}_2.$
See text for further details.
\label{fig1}}
\end{figure}

Inserting the weights  $p_\pm$ and the Bloch vectors
$\vec{r}_\pm$ into  $Q$ {as given by Eq.\ (\ref{Qqubit})} yields 
\begin{equation}
\label{Qpure}
Q= \frac{1}{4} 
\left( 1+\frac{\left(\vec{r}\cdot\Delta \vec{r}\right)^2}
{1-(\vec{r}\cdot\vec{r}_b)^2}\right)\,, 
\end{equation}
Without loosing generality we assume that $\rho_1$ and  $\rho_2$ are  
represented by the Bloch vectors $\vec{r}_1=(0,0,1)$ and 
$\vec{r}_2=(\sin\theta,0,\cos\theta)$, respectively. 
In this case $\vec{r}_b=(\vec{r}_1+\vec{r}_2)/2=
\left(\sin\theta\vec{e}_x+(1+\cos\theta)\vec{e}_z\right)/2$ 
and
$\Delta\vec{r}=
\left(-\sin\theta\vec{e}_x+(1-\cos\theta)\vec{e}_z\right)/6$ 
are perpendicular, see Fig.~\ref{fig1}.
Because of this orthogonality $Q$ in \refe{Qpure} reduces to
\begin{equation}
  Q= \frac{1}{4} \left( 1+\frac{\left(\Delta r\right)^2\cos^2\alpha}{1-r_b^2\sin^2\alpha}\right)\,, 
\end{equation}
where $\alpha$ is the angle between $\vec{r}$  and $\Delta\vec{r}$,
$\Delta r=\|\Delta\vec{r}\|$ and $r_b=\|\vec{r}_b\| \leq 1$. 
$Q$ assumes its maximum value 
\begin{equation}\label{Qmax}
Q_{\mbox{\small max}}= \frac{1}{4} \left( 1+\left(\Delta r\right)^2\right)
\end{equation}
at $\alpha=k\pi$ with $k=0,\, 1,\,2,\ldots$, i.e., 
if the Bloch vectors
$r_\pm$ of the projectors {point into the same (opposite) direction 
as} $\Delta \vec{r}$. This is the case for measurements of 
non-degenerated observables $O$ which commute with
$\rho_1-\rho_2$. 
  
It remains to show, that this von Neumann measurement is also 
optimal among all possible POVM. This we will do in the next 
section, where we 
solve the problem for general $\rho_1, \rho_2.$

\subsection{Mixtures of mixed states}

Which is the optimal measurement, if $\rho_1$ and   $\rho_2$ 
are mixed qubit states? {In order} to answer this question, we first 
determine the optimal projector-valued measure and then show 
that this leads also to a maximal value of $Q$ for all POVMs. 
According to Eq.~(\ref{Qpure}), $Q$ for a given PVM can be 
written as 
\begin{equation}
\label{QPVM}
Q (\alpha) = \frac{1}{4} 
\left( 1+\frac{\left(\Delta r\right)^2\cos^2(\alpha)}
{1-r_b^2\cos^2\beta}\right)\,, 
\end{equation}
where $\alpha$ is the angle between the Bloch vector $\vec{r}$ which
represents one projector of the PVM and  $\Delta\vec{r}$ and $\beta$ 
is the angle between 
$\vec{r}$ and $\vec{r}_b$. Introducing the angle $\gamma$ between
$\Delta\vec{r}$ and $\vec{r}_b$ and taking into account that
$\beta=\alpha+\gamma$, one can directly verify that $Q$ assumes its 
maximum at 
\begin{align}
\label{opt2}
\alpha_0=&
\pm
\arccos\left[
  \frac{\cos\gamma}{\sqrt{\frac{r_b^2}{2}(r_b^2-2)(1-\cos(2\gamma))+1}}\right]\nonumber
  \\
&- \gamma +
  k\pi\quad\mbox{with}\quad k=0,1,\ldots\,.   
\end{align}
where the positive sign is for $0\leq\gamma<\pi,$ and the 
negative sign has to be taken for $-\pi\leq \gamma <0$ 
(as it is in Fig.~2). If both states were pure (as in the 
previous section), we had $\gamma =\pi/2$ and hence 
$\alpha_0 = k \pi.$ For mixed states, however, this is 
in general not the case, hence the projectors of the 
optimal PVM-measurement do not correspond to the eigenvectors 
of $\rho_1-\rho_2$ (see Fig.~2). 
\begin{figure}
\psfrag{r1}{$\vec{r}_1$}
\psfrag{r2}{$\vec{r}_2$}
\psfrag{r12}{$\vec{r}_1-\vec{r}_2$}
\psfrag{ra}{$\vec{r}_a$}
\psfrag{rb}{$\vec{r}_b$}
\psfrag{rr}{$\vec{r}$}
\psfrag{dr}{$\Delta\vec{r}$}
  \begin{center}
   \epsfig{file=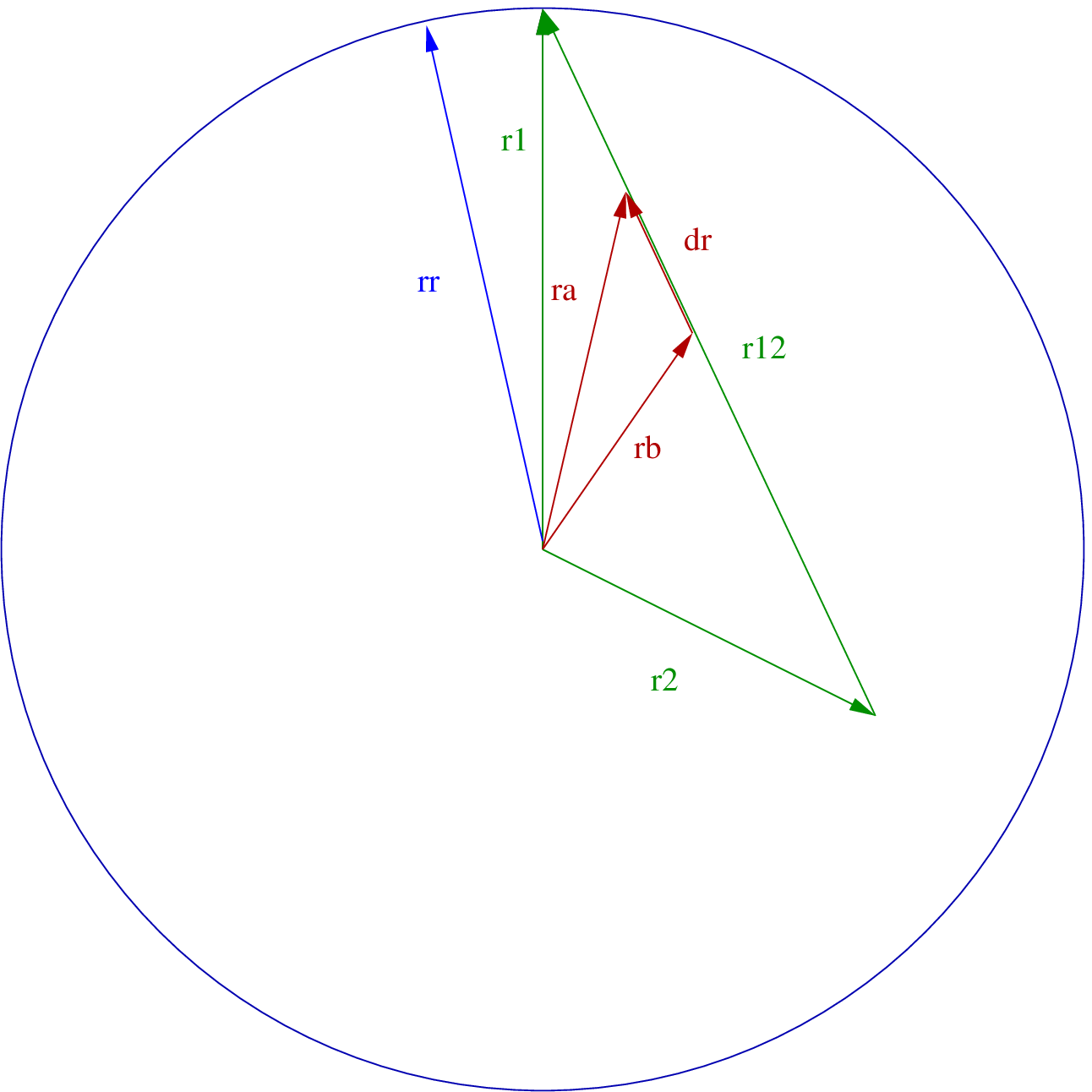,width=0.8\linewidth}
  \end{center}
\caption{Schematic figure of the Bloch representation for 
the case where $\rho_1$ is pure, but $\rho_2$ not. Now, the optimal measurement
is a measurement in the direction of  $\vec{r}$ which is tilted 
towards $\vec{r}_1.$
\label{fig2}}
\end{figure}
Only if both states $\rho_1$ and $\rho_2$ 
have the same purity, (that is, $\tr(\rho_1^2)=\tr(\rho_2^2)$ 
which implies $r_1=r_2 \Leftrightarrow \gamma=0$) the best 
measurement is again that of a non-trivial observable $O$ 
which commutes with $\rho_1-\rho_2$.

The deviation from the pure state case $\alpha=k\pi$ is shown in Fig.~3.
\begin{figure}
\begin{center}
   \epsfig{file=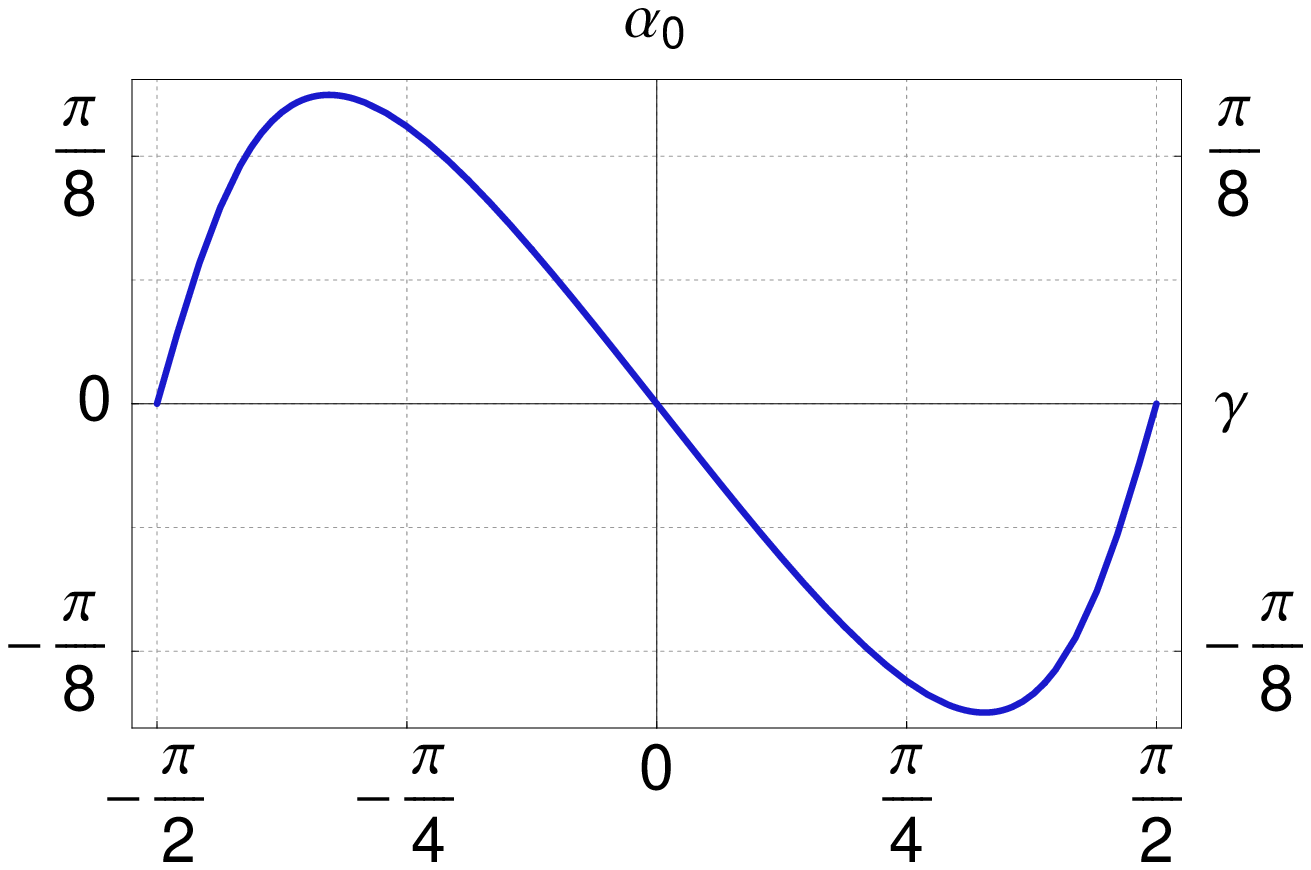,width=0.88\linewidth}
\end{center}
\caption{Deviation of optimal measurement direction 
$\vec{r}$ from direction of
  $\vec{r}_1-\vec{r}_2$ specified by the angle $\alpha_0$ as
  function of angle $\gamma$ between $\Delta\vec{r}$ and
  $\vec{r}_b$ for $r_b=0.8$. For states $\rho_1,\,\rho_2$ 
with the same entropy
  ($\gamma=\pm \pi/2$) the deviation is zero. If, say, state $\rho_1$
  corresponds to less entropy $\vec{r}$ is tilted towards the Bloch
  vector of $\rho_1$. For example compare the case depicted in
  Fig.~\ref{fig2}, where  $\pi/2<\gamma<0 $ and thus $0<\alpha_0<\pi/2$,
  indicating the named tilting.      
\label{fig3}}
\end{figure}
Physically, this deviation may be explained as follows. 
If $\rho_2$ is relatively mixed compared to $\rho_1,$ then
{no  observable does} efficiently resolve the value of 
$\lambda$ for small $\lambda.$ This is simply due to the 
fact that the possible states with a small $\lambda$ are 
close to the maximally mixed state. So it is favorable to 
measure in the direction of $\rho_1$ which gives at least 
a high resolution for high values of $\lambda.$

Finally, we have to show that the optimal projection measurement 
is also the optimal POVM measurement. To do so, 
we prove that the $Q$-value for all POVMs
is less or equal than its  maximum value $Q_{\mbox{\tiny max}}^{\mbox{\tiny PVM}}$ 
for PVMs. For a general Qubit-POVM, (see Eq.~(\ref{Qqubit})), $Q$ is given 
by
\begin{equation}
Q  = \frac{1}{4} 
\left( 1+\sum_{m}p_m\frac{\left(\Delta r\right)^2\cos^2\alpha_m}{1+r_b\cos(\alpha_m+\gamma)}
\right)\,,
\end{equation}
whereas for the maximum $Q$ for projection measurements is of the form:
\begin{equation}
Q_{\mbox{\tiny max}}^{\mbox{\tiny PVM}}= \frac{1}{4} \left( 1+\frac{\left(\Delta r\right)^2\cos^2\alpha_0}{1-r_b^2\cos^2(\alpha_0+\gamma)}\right)\, 
\end{equation}
with $\alpha_0$ given by Eq.~(\ref{opt2}). We can formulate: 
\\
{\bf Proposition 4.} For qubits POVMs can not improve the optimal projective 
measurement, with $\alpha_0$ given in Eq.~(\ref{opt2}). That is, we have
for all POVMs $Q\le Q_{\mbox{\tiny max}}^{\mbox{\tiny PVM}}$.

{\it Proof.} We introduce  the positive function 
\begin{equation}
\alpha\rightarrow  f(\alpha)=\frac{\cos^2(\alpha)}{1-r_b^2\cos^2(\alpha+\gamma)},
\end{equation} 
then the claim can be expressed  as follows:
\begin{eqnarray}
 Q \le Q_{\mbox{\tiny max}}^{\mbox{\tiny PVM}}
&\Leftrightarrow &  \sum_{m}p_m f(\alpha_m)(1-r_b\cos\beta_m) \le f(\alpha_0) \nonumber \\ 
&\Leftrightarrow &\sum_{m}p_m \frac{f(\alpha_m)}{f(\alpha_0)}(1-r_b\cos\beta_m)
\le 1\,,
\end{eqnarray}
where $\beta_m:=\alpha_m+\gamma$.
Since $f(\alpha)\le f(\alpha_0)$ for all $\alpha\in [0,2\pi]$, it
follows that
 \begin{equation}
\sum_{m}p_m \frac{f(\alpha_m)}{f(\alpha_0)}(1-r_b\cos\beta_m)
\le \sum_{m}p_m (1-r_b\cos\beta_m)=1\,,
\end{equation}
which proves the claim. The right-hand side is equal to unity, because
of the constraints on weights $p_m$ and the  Bloch vectors
$\vec{r}_m$ of the effects (\ref{constraints}), i.e., $\sum_m p_m=1$ and $\sum_m p_m
\cos\beta_m= 0$. 
$\hfill \Box$
  
\subsection{Decoherence as an example}

As an example, let us discuss the following problem. We consider 
a two-level atom under the influence of decoherence. The task is, to 
estimate the decay rate from one copy of the state. The evolution
of the density matrix may be described by the master equation 
\begin{equation}
\frac{\partial}{\partial t} \rho = \frac{1}{i \hbar} [H,\rho] + \mathcal{L} \rho,
\end{equation}
where $H=\hbar \omega \sigma_z /2$ is the Hamiltonian of the atom,
and the incoherent evolution is of the Lindblad form \cite{briegel}
\begin{eqnarray}
\mathcal{L}\rho 
&=& -\frac{B}{2} (1-s) 
\big(\sigma^+ \sigma^- \rho + \rho \sigma^+ \sigma^- - 2 \sigma^- \rho \sigma^+ \big)
\nonumber \\
&& -\frac{B}{2} s 
\big(\sigma^- \sigma^+ \rho + \rho \sigma^- \sigma^+ - 2 \sigma^+ \rho \sigma^- \big)
\nonumber \\
&& - \frac{2C-B}{4}\big(\rho - \sigma_z \rho \sigma_z \big).
\end{eqnarray} 
Here, $s= \bra{0}\rho_\infty\ket{0}$ denotes the population of the exited 
state in thermal equilibrium, $\sigma^\pm = \sigma_x \pm i \sigma_y,$ and
$B$ and $C$ are the decay rates of the expectation values of $\sigma_z$
and $\sigma^\pm.$ 

It can be straightforwardly shown that for the case $C=B$ the time dependent 
density matrix is 
given by
\begin{eqnarray}
\rho(t) &=& U \tilde \rho(t) U^\dagger
\nonumber
\\
\tilde \rho(t) &=& e^{-B t} \rho(0) + (1-e^{-B t})
\begin{pmatrix}
s & 0 \\
0 & 1-s 
\end{pmatrix}
\end{eqnarray} 
with $U=e^{-iHt/\hbar}.$ Note that this form is independent 
from the starting density matrix $\rho(0).$ Now we can ask: 
assuming we know $s,t$ and $\rho(0),$ how can we estimate 
the decay parameter $B$ from a single copy of $\rho(t)$?

Let us assume that we know that at least $B \in [0,B^{\rm max}].$ 
Then the state under consideration is, in the rotating frame,
of the form
\begin{equation}
\rho(\lambda)= \lambda \rho_1 + (1-\lambda) \rho_2, \;\;\;\; 
\lambda = e^{-B t} \in [e^{- t B^{\rm max}}, 1].
\end{equation}
While $B \in [0,B^{\rm max}]$ is equidistributed (i.e., $p(B)=1/B^{\rm max}$ for all $B$), 
the parameter $\lambda$ is not. Its probability density is given by
\begin{equation}
q(\lambda)=\frac{1}{\lambda t B^{\rm max}},
\end{equation}
as can be checked by direct calculation. {Employing the density
$q(\lambda)$ we can calculate the mean variance using Eq.~(\ref{varigen}):}
\begin{align}
\sum_m p(m) &\mbox{Var}_m(\lambda) = \frac{1-e^{-2 t B^{\rm max}}}{2 t B^{\rm max}}-
\nonumber
\\
&- x^2 \sum_m 
\frac{
\big( \tr
\big[
E_m \{ (1- \frac{x t B^{\rm max}}{2})\rho_1 + (\frac{x t B^{\rm max}}{2})\rho_2 \}
\big]\big)^2
}
{
\tr
\big[
E_m \{ x\rho_1 + (1-x)\rho_2 \}
\big]
};
\nonumber 
\\
x&=\frac{1-e^{-t B^{\rm max}}}{t B^{\rm max}}.
\end{align}
The remaining optimization problem is essentially the 
same as in the previous case. The only difference is that 
the weights of $\rho_1$ and $\rho_2$ in the previous case 
[$(2/3, 1/3)$ and $(1/2, 1/2)$] are now replaced by more 
complicated expressions. Due to that, the discussion from 
above for the case where $\rho_1$ and $\rho_2$ are pure, is no longer
valid, since the orthogonality of $\vec{r}_b$ and 
$\Delta \vec{r}$ is not guaranteed anymore. However, the solution
of the general case is directly applicable, the only difference 
is the new definition of $\vec{r}_a$ and $\vec{r}_b.$ Hence,
the calculations of Sec. IV can be applied, and finally
Eq.~(\ref{opt2}) solves the problem of the estimation of 
decay parameter.

\section{Optimal measurement for higher dimensional systems}
Let us now discuss the case of higher dimensional systems. That is, 
we consider {a single copy} of the state 
$\rho_\lambda =\lambda \rho_1+ (1-\lambda)\rho_2$ where the $\rho_i$
are density matrices acting on   a $d$-dimensional complex Hilbert 
space ${\mathcal H}_d.$ 

A general solution for this case is quite complicated. 
However, for many 
important cases the problem can be solved as follows. 
Let us choose $d^2$
operators $G_i, i=0,...,d^2-1$ such that they form an 
orthonormal basis of the operator
space. This means, they are hermitian and fulfill 
$\tr(G_i G_j)= \delta_{ij}.$
We can choose them in {such a way that 
that $\rho_1$ as well as $\rho_2$ can be written as 
linear combinations of
$G_0= \eins/\sqrt{d}, G_1$ and $G_2.$} Any {hermitian} 
operator with trace one can be written as
\begin{equation}
O= \frac{1}{d}\Big( \eins + \sqrt{d^2-d} \sum_{i=1}^{d^2-1} r_i G_i \Big)
\label{higher}
\end{equation}
and if $O$ denotes a pure state we have  $Tr(O^2)=1$ which 
is equivalent to $\sum_i r_i^2 =1.$ 

Now one can apply the results of the qubit 
case: First, one can argue as in Proposition 3 
that the effects of the optimal POVM are linear 
combinations of $G_0, G_1$ and $G_2.$ Then, 
the previous section allows to compute the optimal 
$\alpha_0$ and the corresponding vectors. Note that the 
normalizations $1/d$ and $\sqrt{d^2-d}$ in 
Eq.~(\ref{higher}) are chosen in such a way, that all 
the formulae of the qubit case can be applied without 
modification.

The drawback of this ansatz is that the obtained 
optimal vectors are not 
guaranteed to correspond to valid POVMs. Namely, 
in contrast to qubits, for higher 
dimensional systems the condition $\sum_i r_i^2 =1$ does only imply
that $\tr(O^2)=1,$ but not that $O$ is positive. The conditions for 
positivity for Bloch vectors in higher dimensional systems are quite 
{involved} \cite{kimura, khaneja}. However, if the 
resulting solution
yields positive effects, the obtained solution is clearly the optimal 
one. We will discuss now important examples when this is the case. 

Let us first consider the case when $\rho_1$ and $\rho_2$ have support 
on the same two-dimensional subspace of ${\mathcal H}_d.$ This is for 
instance the  case if $\rho_1= \pro{\psi_1}$ and $\rho_2= \pro{\psi_2}$ 
are pure states. Then, by choosing an appropriate basis of the 
two-dimensional subspace, we can assume that $\rho_1$ and $\rho_2$ are real.
Hence, $G_1,G_2$ can be chosen as the Pauli matrices on the subspace, 
and the solution of the two qubit case can directly be applied. 
{Moreover}, if $\rho_1= \pro{\psi_1}$ and $\rho_2= \pro{\psi_2}$ 
the optimal measurement consists of a von Neumann measurement of 
an observable commuting with $\pro{\psi_1}-\pro{\psi_2}$. 

The other important case occurs if the first state is a pure state, 
$\rho_1=\pro{\psi},$ mixed with white noise $\rho_2= \eins/d$. Then we 
have to choose $G_1 =(\eins - d \pro{\psi})/\sqrt{d^2-1}$ and $G_2$ 
arbitrary.  The optimal measurement is then a von 
Neumann-L\"uders measurement with two effects: 
$P_1= \pro{\psi}$ and $P_2= \eins-\pro{\psi}.$

\subsection{Verifying the production of entangled states}

The last example is similar to a task often occurring in experiments.
Namely, one aims to produce an entangled state $\ket{\psi}$,  however 
noise is added during preparation process. This situation may be modeled
by  writing the actual prepared state as
\begin{equation}
\rho(\lambda)=\lambda\pro{\psi}+(1-\lambda)\frac{1}{d}\eins \,.
\end{equation}
Now one would like to know whether the state is entangled or not.
Our results deliver now a possible strategy for this decision: 
one may estimate $\lambda$ with the methods outlined above, 
and then apply separability criteria to the state 
$\rho(\lambda_{\rm est}).$ However, since only one copy is available, 
this does not allow to detect the entanglement unambiguously. 
It is interesting to see that this method is different from the 
standard method for many copies.  Then, entanglement witnesses 
allow the unambiguous detection, 
since for them a negative mean value guarantees entanglement 
\cite{horo,terhal} 
Indeed, these are different observables: Taking the two-qubit case
and $\ket{\psi}= \alpha \ket{01} + \beta \ket{10}$ the optimal 
witness 
is 
$\mathcal{W}=\pro{00}+\ket{01}\bra{10}+\ket{10}\bra{01}+\pro{11}$ 
\cite{jmo}
which is different from the observable which leads 
to the best estimate 
of $\lambda.$

\subsection{Optimal measurement for commuting states}

Another case where we can determine the optimal measurement to
estimate the qudit state $\rho_\lambda$  in  (\ref{rholambda}) 
is given when $\rho_1$ and $\rho_2$ are commuting states, 
\begin{equation}
[\rho_1,\rho_2]=0\,.
\end{equation}
It turns out that measurements leading to the least mean variance  
are  von Neumann  measurements of observables $O$ which commute 
with $\rho_1$ and $\rho_2$, that is, they have the same eigenvectors
as  $\rho_1, \rho_2$. The proof can be accomplished in two steps. 
First we prove that we can restrict the search for optimal 
measurements  to the class of POVMs with effects commuting 
with  $\rho_1$ and $\rho_2$, then we argue that among these 
measurements the projection measurements yield the best estimation 
of $\rho_\lambda$.
\\
{\bf Proposition 5.} Let $\rho_1$ and $\rho_2$ be commuting, 
$[\rho_1,\rho_2]=0$ and let the POVM $\PP=\{E_m\}$ represent 
a measurement to estimate the state 
$\rho_\lambda=\lambda\rho_1+(1-\lambda)\rho_2$. 
Then there is a projective measurement 
$\tilde{\PP}=\{F_m\}$ with $[F_m,\rho_1]=0$ for all $m$ 
which satisfies $Q(\PP)=Q(\tilde{\PP})$.

{\it Proof.} The  commuting states $\rho_1$
and $\rho_2$ are simultaneously diagonalizable,
\begin{equation}
 \rho_1=\sum_i t_i \pro{i}\quad\mbox{and}\quad
\rho_2= \sum_i s_i \pro{i}.
\end{equation}
Thus $Q(\PP)$, see Eq.~(\ref{QvonP}), can be expressed as
\begin{equation}
Q(\tilde{\PP})=\frac{1}{4}\sum_{m}
\frac{\left[\sum_i e_{mi}\left(\frac{2}{3}t_i+
\frac{1}{3}s_i\right)\right]^2}{\sum_i e_{mi}
\frac{1}{2}\left(\rho_1+ \rho_2\right)} 
\end{equation}
with $e_{mi}:=\bra{i}E_m\ket{i}$ Now, the positive operators 
$F_m:=\sum_i e_{mi}\pro{i}$ form a POVM, i.e., they satisfy 
the completeness relation:
\begin{equation}
  \sum_m F_m = \sum_{m,i}\pro{i}E_m\ket{i}\bra{i} = \eins
\end{equation}
It is easy to see that the commutative POVM 
$\tilde{\PP}=\{F_m\}$  leads to the same
value of $Q$ as $\PP$. Hence, we have only to consider POVMs which
commute with $\rho_1$ and $\rho_2$ to find a measurement which
maximizes $Q$. In addition we already learned  in section
\ref{condopt} that an   optimal POVM is distinguished by pure
effects. Together with the commutativity of the effects it follows
that the optimal measurement is a projective one.  
$\hfill \Box$.

\section{Conclusion}

In conclusion, we have studied  parameter estimation for quantum states, 
when only one copy of the state is available.
For one qubit, we solved the problem by explicitly constructing the 
measurement that minimizes the deviation between the  true value of 
the parameter and the estimated one, using a Bayesian estimator.
Furthermore, we showed how the results from the qubit case can 
readily be used to  solve this problem  for important higher 
dimensional cases. 

Our work can be extended into several directions. First, one may 
look at higher dimensional systems, trying to find general solutions
for this case. Here, it would be of great interest to find cases where, 
contrary to the qubit case, general POVMs allow for a better parameter 
estimation than von Neumann measurements. For practical purposes, it may 
also be relevant to develop optimal measurement strategies for several,
 but a finite number of copies.

\section{Acknowledgment}

{We thank D. Janzing and A. Scherer for useful discussions. 
This work was supported  by the FWF and the EU (OLAQUI, SCALA, 
QICS) as well as the Center for Applied Photonics in Konstanz, 
Germany.}

\end{document}